# Swimming against the Tide: Gender Bias in the Physics Classroom[1]

Amy L. Graves
Dept. of Physics and Astronomy, Swarthmore College
500 College Ave., Swarthmore, PA 19081
abug1@swarthmore.edu

Etsuko Hoshino-Browne
Dept. of Psychology, Neumann University
One Neumann Drive, Aston, PA 19014
hoshinoe@neumann.edu

Kristine P.H. Lui
Science, Engineering and Technology Dept., Montgomery College
20220 Observation Drive, Germantown, MD 20876

## ABSTRACT

This study examines physics students' evaluations of identical, video-recorded lectures performed by female and male actors playing the role of professors. The results indicate that evaluations by male students show statistically significant overall biases with male professors rated more positively than female professors. Female students tended to be egalitarian, except in two areas. Female students evaluated female professors' interpersonal/communicative skills more positively than male professors'. They evaluated female professors' scientific knowledge and skills less positively than that of male professors just as male students did. These findings are relevant to two areas of research on bias in evaluation: rater-ratee similarity bias and stereotype confirmation bias. Results from this study have important implications for efforts focused on educating students and mentoring faculty members in order to increase the representation of women in the physical sciences.

## 1. INTRODUCTION

### 1.1 Underrepresentation of Women in STEM

Although there are STEM fields with gender parity or pro-female bias in the awarding of Ph.D.'s and rates of faculty promotion, more often these fields show a gender gap favoring men in all career stages (NSF, 2014). Women occupy only 9% to 16% of tenure-track positions in academic institutions (Ceci and Williams, 2011; Nelson, 2007). In the case of physics, NSF statistics indicate that over 85% of physicists with doctoral degree in the US are males and over 70% are white (NSF, 2014; Whitten, 2012). Physical sciences is among the most disproportionately gendered of areas (Ivie et al., 2013; Yost et al., 2013), and several recent international conferences of women have shown that gender equality – measured by equity in numbers of participants, resources and job satisfaction – is far from the norm in virtually all countries (AIP, 2013). Despite recent gains in the percentage of women at many levels (Ivie et al., 2013), women are still strongly underrepresented as physics professors, and are expected to remain underrepresented for many decades to come (APS, 2007; Georgi, 2004).

Isolation is one of the five biases recently listed by Williams, Phillips, and Hall (2015) as pushing women of color out of STEM professions. Thus isolation is both a symptom and a cause of underrepresentation. With the tiny numbers of women in physical sciences, professional isolation is a problem for both majority and minority women. Psychological studies (e.g., Murphy, Steele, and Gross, 2007) indicate that women feel a greater sense of belonging in a STEM setting where the population is gender-balanced. In physics, combatting isolation with horizontal mentoring networks and virtual communities has been both the subject of research and ongoing funding by the National Science Foundation (Blaha et al., 2010; Cunningham et al., 2015).

1.2 Repercussions of Underrepresentation

Gender diversity in physics is important from a social-justice point of view and aligns with Step 7 of Whitten's (2012) "(Baby) steps toward a feminist physics." Diversity is also important as a marker of the quality of knowledge generated within the STEM communities (Harding, 2001). In academic institutions, diversity ensures that professors may serve as role models for the next generation, a crucial element in retaining underrepresented students (e.g., Carlone and Johnson, 2007; Drury, Siy, and Cheryan, 2011; Hughes, 2012; Kosoko-Lasaki, Sonnino, and Voytko, 2006). Though survey data (Downing, Crosby, and Blake-Beard, 2005; Hazari et al., 2013) show that role models need not be women in order to encourage interest in women to pursue STEM, once in the discipline, identification with a female role model can "inoculate" one against stereotype threat (Manke and Cohen, 2011; Marx and Roman, 2002; Stout et al., 2011). While identification with a role model can be more subtle than sharing a race or gender (Cheryan et al., 2011), and overly-impressive role models have even been observed to deflate self-concept (Hoyt and Simon , 2011; Lockwood and Kunda, 2000),  there is strong support for the notion that self-similar, supportive role models improve underrepresented students' performance, self-concept, and persistence in STEM fields (Blake-Beard et al., 2011; Drury et al., 2011; Johnson, 2007; Lockwood, 2006; Manke and Cohen, 2011; Newman, 2011; Stout et al., 2011).  For example, the landmark study by Carell et al. (2009) on roughly 9500 US Air force academy undergraduates showed that while professor gender did not impact male students, having a female professor had a dramatic effect on high-achieving women students. Their performance was enhanced in current and future STEM classes, and the likelihood increased for their pursuit of a STEM major.

1.3 Possible Causes of Underrepresentation

Isolation is just one of many socio-cultural and psychological factors that have been blamed for women's lower levels of job satisfaction and greater tendency than men to leave STEM fields (Carrell et al*.,* 2009; Ceci and Williams, 2011; Steele, Spencer, and Aronson, 2002; Valian, 1998). Physics departments in particular are challenging places for women, replete with multiple personal and professional barriers (Rolin and Vainio, 2011). The factor which motivates the current study is bias in the evaluation of women. Studies of implicit bias and automatic stereotyping (e.g., Fazio et al*.*, 1995; Greenwald, McGhee, and Schwartz, 1998; Reuben, Sapienza, and Zingales, 2014; Sinclair and Kunda, 2000; Wilson, Lindsey, and Schooler*,* 2000) suggest that women routinely encounter biased evaluations which lead to discrimination in hiring and promotion. Bias is apparent, for example, in recommendation letters (Madera, Hebl, and Martin, 2009; Trix and Psenka, 2003) and in ratings that impact hiring decisions (Davison and Burke, 2000; Johnson, Hekman and Chan, 2016; Moss-Racusin et al*.,* 2012; Reuben et al*.*, 2014).

Recruitment and retention of women depends on the evaluation process, thus it is vital to recognize the existence of bias in the evaluation of a woman professional in STEM. Negatively biased evaluations harm careers in several ways; they not only act as an obstacle to professional advancement, but foster a low self-concept (a gloomy picture of one's own strengths and status) and low job satisfaction. The importance of being recognized as a capable "science person" was shown to influence persistence in research on women of color (Carlone and Johnson, 2007). An important climate issue is that of being a good "fit" within an institution (Gallagher and Trower, 2009; Yost et al., 2013). In the academy, being fit for one's job is measured to a great extent by student evaluations (Aleamoni, 1999; Benton and Cashin, 2014), the subject of the current study.

1.4 A rationale for gender bias

Explicit stereotyping is a controlled process, in which a person consciously uses stereotypes based on the target's group membership and category, such as "women". On the other hand, implicit stereotyping or bias is an automatic process, which does not rise to the conscious level. In this second decade of this millennium, whether explicit or implicit bias impedes women is a difficult question, and very much tied to the nature of a profession and the local culture in which it is embedded. As such, despite the fact that (in most countries of the world) explicit bias in the workplace is outlawed, individuals in "stereotype-incongruent" occupations (e.g., female physicists or male kindergarten teachers) can still feel the effects of bias. Experimental evidence for the existence of implicit bias is extensive (Brescoll, Dawson, and Uhlmann, 2010; Greenwald and Krieger, 2006; Institute of Medicine, National Academy of Sciences, and National Academy of Engineering, 2007; Lemm and Banaji, 1999; Valian, 1998). Stereotypes describe women as communal and deficient in the agentic male-typed traits associated with competence and successful performance in fields historically occupied by men (Eagly and Karau, 2002; Valian, 1998). Consequently, women are more easily rated as incompetent, held to higher standards to confirm competence than men in these fields, yet penalized for agency-requiring success. Women are also uniquely punished for certain behaviors necessary in the professional world, such as giving negative feedback (Sinclair and Kunda, 2000) or being perceived as less-than-agreeable (Heilman et al., 2004; Rudman and Glick, 1999).

Thus, stereotypes lead to women needing to out-perform men to be rated as equally competent, yet gender stereotypes also create expectations that women embody communal traits. The assertive, even aggressive (Georgi, 2004) behavior of a confident physicist must somehow be reconciled with behavior that is modest and communal (Basow, 1995; MIT, 2011). Women must walk a 'tightrope' (Williams, 2015) in order to be deemed competent in a field like physics.

1.5 Bias in Student Evaluations

A typical college course evaluation is summative, and is often utilized by administrators to determine a faculty member's worth, and to decide whether to retain and promote (Aleamoni, 1999; Benton and Cashin, 2014; Marsh, 2007). Unfortunately, there is debate over whether such instruments necessarily correlate with evidence-based assessment of student gains, and whether they predict student performance on tests and/or in subsequent classes (Braga, Paccagnella, Pellizzari, 2011; Carpenter, 2013; Carrell, 2009). Further, traditional evaluations *do* correlate with characteristics (personal appearance, style of delivery) that an administrator would view as spurious to instructor worth (Basow, 1998; Joye and Wilson, 2015; Stark and Freishtat, 2014; Neath, 1996). A partial list of performance-related items which may skew student evaluations are the perceptions of scientific competence (Moss-Racusin et al, 2012), perception of mathematical ability (Reuben et al, 2014), agency (Trix and Psenka, 2003), demeanor (Heilman et al, 2004), emotional warmth (Linse, 2003), and physical attractiveness (Hamermesh and Parker, 2005).

If faculty are held accountable to gendered expectations by students, with a larger and often inconsistent set of expectations placed on women (Anderson and Smith, 2005; Basow, 1998; Bennet, 1982; Sinclair and Kunda, 2000; Sprague and Massoni, 2005), then instructor gender may influence teaching effectiveness as seen through the lens of teaching evaluations (Kaschak, 1978; Reid, 2010). For example, in the recent study by MacNell, Driscoll, and Hunt (2015), it was possible to switch the perceived genders of one male and one female professor in an online social science course. A perceived male identity netted the professor higher scores on promptness, fairness, enthusiasm, and other hallmarks of good teaching. To complicate the picture further, interactions which affect ratings have been noted between race and gender of the professor (Ho, Thomsen, and Sidanius, 2009; Hamermesh and Parker, 2005; Huston, 2005; Reid, 2010) and/or the student (Anderson and Smith, 2005; Basow and Silberg, 1987; Bavishi, Medera, and Hebl, 2010; Sprinkle, 2008).

The literature on gendered evaluation of professors suggests that women in STEM fields would be likely to receive biased evaluations from students. Huston (2005, 2006) finds that bias against female faculty is particularly likely in male-typed disciplines, and may be attributed to either explicit or implicit gender stereotyping (Devine, 1989; Greenwald and Banaji, 1995; Wilson et al., 2000). Direct support for gender bias in student evaluations of STEM faculty can be found in literature reviews (Basow and Martin, 2013; Huston, 2005; Martin, 2015). There is also evidence that gender of the student evaluator and STEM instructor will interact (Basow and Silberg, 1987). One important study, albeit for teachers the high school level (Potvin et al., 2009) found that male college students gave lower ratings to their former women teachers in biology, chemistry and physics, while female students gave lower ratings only to the women physics teachers. (Teachers of both sexes were equally effective in terms of students' performance outcomes.) In the literature on gender bias in college teaching evaluations, STEM has unfortunately received comparatively little scrutiny, which has been attributed to the difficulty of obtaining good statistics from small numbers of students (Basow, 1995, 1998, 2011).

In light of our current study, three previous studies should be mentioned. These involved evaluation of identical lectures, but employed abstractions rather than human lecturers. Using a stick figure with stated gender (male or female) and age (young or old) and a neutral lecturing voice, Arbuckle and Williams (2003) showed that gender did not correlate with items such as organization, precision, or use of scientific terms. However, attributes like enthusiasm were significantly different, with young male "professor" being rated as most enthusiastic and interested in student learning and success. In the study by Basow, Codos, and Martin (2013), a computer-animated "talking head" (White or African American, and male or female) was voice-modulated according to gender. The "unnaturalness" of the instructor was evaluated, and found to be significant. Evaluations of this simulated instructor delivering a lecture in engineering yielded no main effect of professor gender or significant interaction between professor and student gender. In Joye and Wilson (2015), pictures of a male and female instructors were selected, and digitally aged to produce young and old conditions of each. The audio was contributed by one person, whose voice was digitally altered to be gender and age ambiguous. This study found that younger professors were rated as having greater rapport with students, for both professor genders; and male professors were rated as more effective, for both professor ages. Interestingly, despite her significantly lower effectiveness and rapport scores, the older female produced the highest quiz scores among students – a result that the authors attribute to increased focus, perhaps in order to please what they perceived as a mother archetype.

1.6 The Current Study

We investigated whether professor gender influences student perceptions of teaching effectiveness. Our testing instrument was a scripted and digitally prerecorded lecture – a simulated classroom situation, with actors portraying physics professors.[4] The recording included technical content appropriate for introductory physics students, chalkboard calculations, and a demonstration involving a laser. The scripted lecture had a built in "mistake" which the professor noticed and corrected, and the professor provided answers to a couple of questions from the class (all scripted). We wrote and administered summative teaching evaluation, in order to probe for systematic biases attributed to the gender of the lecturer. Female and male students judged a professor for her or his knowledge of physics, teaching competence, and caliber as a job candidate.

---

[4] While a potentially interesting gender dimension exists between being called an "instructor" as opposed to a "professor," this dimension was not explored in our study. The IRB instructions to subjects and the subsequent questionnaire were consistent and always identified the person giving the lecture as a "professor." Note that we use "instructor" and "professor" interchangeably in this text. Further, the more accurate appellation "lecturer" is used in our quantitative methods and results sections, including figures.

Our broad hypothesis was that the females would be rated more poorly than males in their role as a physics professor. Our research questions going into this study were:

• Would students' overall rating of the professor be influenced by student gender, professor gender, interaction between student and professor gender, or institutional differences?

• For subsets of questionnaire items, chosen a priori to probe gender-stereotypes relevant to specific skills, would the professor's gender influence the student ratings?

• Would the recommendations by students to hire the professor, were he/she a job candidate, be influenced by student gender, professor gender, interaction between student and professor gender, or institutional differences?

2. METHODS

2.1 Participants

A total of 126 undergraduate students (47 females and 79 males) from introductory physics classes at two research intensive private universities in the U.S., termed A and B, participated in this study. Their mean age was 19.12 ($SD = 1.24$). University A, which admitted roughly 10% of applicants at the time of the study, is referred to in what follows as being "more selective" than B, which admitted roughly 40% of applicants. The undergraduate populations of A and B are each approximately 5000 students. The setting of A is an affluent suburb in an Eastern state, while B is located in an urban area in the Midwest. Both have strong reputations for excellence in educating STEM students.

2.2 Rehearsal and production of videotaped lectures

Two female and two male professional actors (Anglo Americans of similar physical attractiveness and age) were rehearsed in order to deliver a scripted lecture in an identical manner. The script of the physics lecture was based on Steven Weinberg's 1986 Dirac Memorial Lecture and Swarthmore College freshman quantum mechanics lectures, particularly those authored by Professor John Boccio. The two female and two male actors viewed recorded and live physics lectures given by male and female professors with acknowledged skill at teaching college-level physics. Then, lines and blocking (i.e. the words and physical actions) were rehearsed in the presence of the first author of this paper, who served as the director. All actors were present at all rehearsals. This protocol reinforced the goal of performing the scripted lecture with good teaching style, and in an identical manner for all actors. Their performances were digitally video-recorded and edited, resulting in four versions of the same lecture. The recorded lecture, approximately eight minutes duration, included speaking, writing on the blackboard, a hands-on demonstration of how laser light is polarized in different crystallographic directions as it passes through a special crystal (Keilich and Zawodny, 1975), questions by "students" (actors voicing the questions off-camera), and the corresponding on-camera answers by the actor playing a professor. Footage of college students attending a real class in the classroom was obtained later and was intercut with the actors' performances in an identical manner in all four video-recorded lectures. The demonstration showing the laser light was treated in a similar manner, so that all four videos contained identical demonstration footage. Finally, before these videos were used in the study, one version of the videos was shown to a focus group of physics students at Swarthmore College to confirm the clarity of lecture content, its credibility as a typical freshmen physics class, and the face validity of the survey instrument.

2.3 Procedure

This research involving human subjects was approved by the Institutional Review Board of Swarthmore College. Participants read and signed a consent form which was approved by the IRB, affirming their willingness to participate and assuring anonymity and confidentiality. Participants were told that the purpose of the study was to investigate factors that influence student evaluations of a physics lecture and of the professor giving the lecture. They were randomly assigned to a room in which they were able to watch one of the four video-recorded lectures. They were not aware that there was more than one version of the lecture being shown to students on their campus. After watching the video-recorded lecture, the participants filled out a questionnaire. After a first question meant to orient participants to the task of evaluating a physics lecture, there were 16 questions that assessed aspects of the either the lecture or the professor's abilities on a 5-point scale (see Appendix A). Questions 2-10 and 12-14 had responses from "disagree strongly" to "agree strongly." Question 11 asked about the amount of math done in the lecture, and responses ranged from "much more than seemed necessary" to "much less …." The last three questions were typical, traditional attempts to summarize overall quality (Hobson and Talbot, 2001). Questions on the lecture (15) and the professor (16) had responses from "poor" to "excellent." (Raters have been known to employ gender-skewed standards of "excellence" (Biernat and Kobrynowicz, 1997); however, our experimental design precluded calibrating expectations, based on knowledge of professor gender.) Our study was designed to probe the impact of instructor gender on real-world course evaluations, and on the critical issue of hiring decisions, which prompted one more summative question (17), asking for a hiring recommendation. At the end of the questionnaire, participants also had an opportunity to share their thoughts in an open-ended response section, where they were invited to comment further on the lecture or the professor.[6]

The questionnaire was designed to be both a valid evaluation of teaching, and one that raised no suspicion in subjects as to the purpose of the study. Questions were consistent with templates set forth by major centers for college teaching and learning (Iowa State Univ., 2015, Berkeley Center for Teaching and Learning, 2015).[7] Further, our questionnaire aligns with Danielson's "Framework for Teaching Evaluation Instrument" of the ASCD (Danielson, 2013).[8]

We probed: organized presentation of content, content at a correct level, presentation at a correct speed, comprehensibility of lesson, effectiveness of teaching to promote learning, content knowledge on the part of the instructor, perception that students seen in the video had good interactions with instructor, perception that the instructor would be approachable for interaction outside of the class, and questions of subjective affect (boredom, enjoyment, likeability of instructor). The evaluation ended with the traditional, general summative questions. Clearly, many questions could have been chosen for inclusion. For the current study, social role theory and previous work on the role of gender in evaluating STEM professionals guided our specific choices. The questions "This professor has a solid grasp of the material being taught" and "The professor is not a very knowledgeable physicist" address presentation of content, but in a way that touched upon the stereotype of males as scientific authorities (Chambers, 1983; Scheibinger, 1999). The question "This professor goes too slowly during a class" addresses teaching practices, but in a way that touches upon gender stereotypes of appropriately high-level teaching (college vs. K-12). The question "This professor is good at handling experimental equipment" touches upon the stereotype of men having superior visuo-spatial abilities (Levy and Kimura, 2009).

---

[6] We hope to publish the open-response analysis in a future paper.

[7] All template categories that could be plausibly addressed by viewing a portion of a class were included, and the only omissions were inapplicable categories (in particular, those that asked about actual involvement in a class.)

[8] Developed for K-12 educators, this document has been revised over three decades based on theory and praxis. Our questionnaire (see Appendix A) contains questions linked to Danielson's domains 1a, 1c-1e, 2a, 2c, and 3a-c.

From the questionnaire items, we also created three a priori composites: scientific thinking and hands-on skills (stereotypically male), interpersonal and communicative skills (stereotypically female) and lecture quality (no gender stereotype assigned; a general summative question which should yield results consistent with the average over all questionnaire items).

Once the questionnaire and the open-ended responses were completed and collected, participants completed a section including demographic items regarding participants' age, sex, class year, whether or not they were native English speakers, and whether they took physics courses in high school.

3. RESULTS

Results reported below were based on 121 participants (42 female, 79 male) after excluding five participants from data analysis who indicated their suspicion that the lecturer was not an actual physics professor. To examine our major research questions probing an interaction between gender of the student and gender of the professor affect student evaluation of the professor, we used sex and lecturer sex as our major independent variables in all analyses. We used the analysis of variance (ANOVA), with a critical alpha of 0.05. The ANOVA was used because it was assumed that ratings would be normally distributed, variance would be relatively homogeneous, and that ratings are independent of one another. Although the analysis of variance is fairly robust to violation of normal distribution of observations and homogeneity of variance, we confirmed that our data were mostly normally distributed and variance was homogeneous. The data were analyzed using SPSS (Statistical Package for Social Sciences) version 19.

3.1 Analysis of the overall evaluation score

Fifteen questionnaire items (excluding the first item which was not directly relevant to the quality of lecture or the capability of professor, and the eleventh, whose two extremes embodied different critiques of the professor's capability) were combined to create an "overall evaluation" score, which was our major dependent variable. This overall evaluation composite was very reliable; Cronbach's alpha was .92. The overall evaluation score is the average of the scores on 15 questions (reversed if necessary so that 5 is most positive, and 1 least positive). To test whether students' overall rating of the professor would be influenced by professor gender, we performed a 2(lecturer sex: female vs. male) x 2(student sex: female vs. male) ANOVA. See Appendix A for item statistics. See Table 1 for detailed statistical results. The results yielded neither a main effect of lecturer sex nor a main effect of student sex. However, the interaction of two variables was significant, $F(1, 117) = 4.91$, $p = .029$ (Table 1 and Figure 1). Female students evaluated female lecturers ($M = 3.73$, $SD = 0.70$, $n = 21$) and male lecturers ($M = 3.55$, $SD = 0.75$, $n = 21$) more or less equally. However, male students evaluated male lecturers significantly more positively ($M = 3.88$, $SD = 0.75$, $n = 39$) than female lecturers ($M = 3.44$, $SD = 0.71$, $n = 40$).

To test whether the results would be sensitive to institutional differences, we performed a 2(lecturer sex: female vs. male) x 2(student sex: female vs. male) x 2(institution: more selective university vs. less selective university) ANOVA. There was a significant main effect of institution, $F(1, 113) = 8.42$, $p = .004$, but this variable did not interact with lecturer sex or student sex. Thus, while the mean of overall evaluation score was significantly lower at the more selective university ($n = 66$) than at the less selective university ($n = 55$), much stronger positive evaluations by male students of male lecturers than female lecturers were found at each university independently (Figure 2).

We also explored the independent variable of lecturer. We performed a 2(student sex: female vs. male) x 4(lecturer: female lecturer 1 vs. female lecturer 2 vs. male lecturer 3 vs. male lecturer 4) ANOVA to examine whether each lecturer was evaluated similarly. Neither a main effect of student sex nor an interaction between student sex and lecturer was significant, $F(1, 113) = 0.08$, $p = .782$ and $F(1, 113) = 2.04$, $p = .112$, respectively. However, a main effect of lecturer was marginally significant, $F(1, 113) = 2.55$, $p = .06$. We examined the extent to which this marginally significant effect of lecturer was consistent across female and male students' ratings by performing simple main effect tests. In order to avoid Type I error from multiple comparisons, we used a lower critical alpha of .01 rather than the conventional .05 for these additional analysis. We found that whereas female students evaluated all four lecturers more or less equally, $F(1, 113) = 0.72$, $p = .402$, male students showed more variation in their evaluation, $F(1, 113) = 4.86$, $p = .027$. Specifically, the male students evaluated one of the male lecturers more positively ($M = 4.20$, $SD = 0.61$, $n = 17$) than one of the female lecturers ($M = 3.32$, $SD =$ 0.72, $n = 20$). However, this comparison was not statistically significant with the lower critical alpha ($p = .01$).

3.2 Analysis of specific skills and lecture quality

Would the professor's gender influence the student ratings of gender stereotype-related items? We analyzed three a priori composites which had been created before raw data were viewed: scientific thinking and hands-on skills; interpersonal and communicative skills; and lecture quality. These composites showed good reliability; Cronbach's alphas indicated .76, .71, and .79 respectively. Each of these composites was submitted to a 2(lecturer sex: female vs. male) x 2(student sex: female vs. male) ANOVA (see Tables 2, 3, and 4 for a list of composite items and detailed statistical results).

For the evaluation of scientific thinking and hands-on skills, there was a significant main effect of lecturer sex, $F(1, 117) = 8.24$, $p = .005$ (Table 2 and Figure 3). Both female and male students were positively biased toward male lecturers by evaluating them to have better scientific thinking and hands-on skills than female lecturers. An exploratory simple main effect test indicated that this bias was much stronger among male students. Neither a main effect of student sex nor an interaction between lecturer sex and student sex was significant.

In contrast, the evaluation of interpersonal and communicative skills showed a same-sex bias (Table 3 and Figure 4). That is, while main effects of lecturer sex and student sex were not significant, these variables interacted significantly, $F(1, 117) = 7.41$, $p = .007$. Simple main effect tests indicated that female students evaluated female lecturers on their interpersonal and communicative skills more positively than male lecturers, whereas male students evaluated male lectures more positively than female lecturers.

The third composite, lecture quality, yielded results which, as hypothesized, were similar to the overall evaluation (Table 4). There was neither a main effect of lecturer sex nor a main effect of student sex. An interaction between the two variables was marginally significant, $F(1, 117) = 3.50, p = .064$. Simple main effect tests indicated that whereas female students evaluated the lecture quality of female and male lecturers more or less equally, male students evaluated the lecture quality of male lecturers more positively than that of female lecturers.

### 3.3 Analysis of the hiring recommendation item

We also examined whether the recommendations by students to hire the professor, were she/he a job candidate, would be influenced by professor gender. A 2(lecturer sex: female vs. male) x 2(student sex: female vs. male) ANOVA yielded a pattern identical to the overall score results. There was neither a main effect of lecturer sex nor a main effect of student sex. However, these two variables interacted significantly, $F(1, 117) = 5.44, p = .021$. Simple main effect tests indicated that female students recommended female lecturers ($M = 3.38, SD = 1.12, n = 21$) and male lecturers ($M = 3.14, SD = 1.11, n = 21$) approximately equally, $F(1, 117) = 0.49, p = .492$. However, male students recommended male lecturers more strongly ($M = 3.69, SD = 1.10, n = 39$) than female lecturers ($M = 2.95, SD = 1.08, n = 40$), and this difference was significant, $F(1, 117) = 8.99, p = .003$.

We also examined the correlation between the rating of this hiring recommendation item and the mean of the rest of the evaluation items. As expected, the correlation was very strong in the positive direction, $r(119) = .84$, $p < .001$. Thus, students who evaluated the professor and her or his lecture more positively also recommended the professor more strongly to be hired at their universities. This strong positive correlation between positive evaluation of the professor and hiring recommendation of the professor was maintained even when female students and male students were analyzed separately: $r(40) = .86$, $p < .001$ and $r(77) = .83$, $p < .001$ respectively.

## 4. DISCUSSION

### 4.1 Results of the current study

We examined male and female students' perceptions of an identical, video-recorded physics lecture performed by two female and two male actors playing the role of professors. The broad hypothesis was *not* upheld. When all student responses were considered, there was no effect of the gender of the lecturer. However, our research questions concerning interaction between gender of student and professor yielded interesting results. Male students judged female physics professors more poorly on these metrics: overall evaluation score, hiring recommendations, hands-on and intellectual skills, and approachability and interpersonal skills. Female students, on the other hand, rated male and female physics professors equally, except in the area of scientific thinking and hands-on skills, which showed bias consistent with the male-as-scientist stereotype; and that of interpersonal skills, in which female students were biased towards female professors. Our results were robust across the two institutions studied. That is, as Figure 2 shows, while the more selective institution yielded significantly lower ratings for all professors regardless of gender, the interaction between student and professor gender remained a significant effect.

Our study provides further empirical evidence for findings mentioned in the Introduction, such as those found in correlational studies of professionals in gender-incongruent occupations, experimental studies in which the names on resumes or scholarly papers were manipulated, studies in which actors (or a single actor, voice modulated) performed identical scripts, and studies in which gender was correlated with results of actual college teaching evaluations. However, the specific design of the current study - a strongly male-typed STEM field, with student gender as a variable - is sufficiently novel that comparison with previous studies is not straightforward. The real-world course evaluation study of Basow (1995) agreed with ours in that there was a gender interaction, with male students rating female professors more negatively than male professors, and more negatively than did female students in the same class. However, an animated "talking head" delivering an engineering lecture (Basow et al., 2013) produced neither a main effect of professor gender nor an interaction between student and professor gender (though quiz results were higher for the "normative" professor). One study (Study 2) of Sinclair and Kunda (2000) employed only male students to rate a video. Thus, their result in which students downgraded female professors who delivered negative feedback resonates with ours. Basow and Silberg (1987) reported on interactions between student and professor gender in real-world course evaluations across a variety of fields. Male students rated female professors less favorably in overall teaching ability, and on all 6 measures studied. Female students also rated female professors more poorly on overall teaching ability, and only 3 measures. However, these results cannot be claimed to be present when evaluations are confined to STEM courses. Kaschak (1976) found that whether a field was gender-congruent did not affect evaluations of a fictitious professor by students. However, as in our study, the professor's "sex … seemed to be the crucial one on which faculty members were evaluated by male students" (Kashak, 1976, p. 241). On all scales save one that measured powerlessness, female students were egalitarian in their evaluations.

Our broad hypothesis that female professors would be rated more poorly than males, regardless of student gender, was not upheld. This contrasts with, for example, studies in which disembodied voices delivered lectures. Two studies involving social sciences were Joye and Wilson (2015) and Arbuckle and Williams (2003). Joye and Wilson found male professors rated as more effective, while Arbuckle and Williams found that the lecturer had to be both young and male to receive significantly better ratings. The online learning study of MacNell et al. (2015) showed that a male perceived identity netted the instructor significantly higher scores on an overall rating, as well as 6 of 12 individual measures of teacher effectiveness. The male instructor further received significantly higher scores on 3 "interpersonal" measures, akin to our a priori measure of three characteristics indicative of interpersonal skills, where our evaluations divided along same-gender lines.

4.2 Limitations of the current study

Admittedly, one should be cautious about the generalizability of our physics-related results to other STEM fields and other institutions. Our research is based on data from freshman students enrolled in physics classes at two academic institutions. To strengthen the generalizability, or to discern distinctions between fields – in particular between physical and life sciences – data from many more academic institutions and from a wide variety of students and STEM classes are needed, given differences in gender ratios within these fields (e.g., CWSEM, 2014; Ginther, 2006; NSF, 2014). Additionally, one might suspect that the female and male actors who played the "professor" somehow conveyed inherently female-related communalism versus male-related agency and competence in their lecture styles, despite using an identical script and delivering an identical lecture. For instance, anything from vocal pitch, speech mannerism, body shape, to subtle facial expressions of the actors might inadvertently conveyed a female communal or male agentic message, even if the words spoken and the physical gestures were made identical between the actors. We acknowledge these criticisms to be valid. However, the current findings are regrettably consistent with past findings by numerous researchers who studied sex discrimination in the natural sciences, as reviewed above.

4.3 Possible theoretical frameworks

In assimilation bias (Biernat and Kobrynowicz, 1997) as well as confirmation bias (Nickerson, 1998), people favor information which support preconceived notions, including racial or gender stereotypes, and draw conclusions that perpetuate the initial beliefs. The tendency of both our male and female participants to evaluate male physics lecturers more favorably than female counterparts on the scientific thinking and hands-on skills is consistent with gender bias and the male-as-scientist stereotype. In such a male-typed field as physics, it may not be a surprise that students form a biased expectation that male professors possess more scientific skill than female professors. On the other hand, a "shifting-standards" framework for how stereotypes influence judgment (Biernat and Kobrynowicz, 1997; Biernat, Fuegen, and Kobrynowicz, 2010) seems less applicable to our results, as it would predict enhanced subjective ratings of female physics professors, who as a group would be held to a lower competency standard.[9]

The biases in student ratings found in our data could be a result of implicit stereotyping. Although we did not directly measure the implicit bias each student held regarding gender and physics, given research findings in implicit prejudice (e.g., Ceci and Williams, 2011; Devine, 1989; Fazio et al., 1995; Greenwald and Banaji, 1995; Lepore and Brown, 1997; Nosek et al., 2009; Rudman, Greenwald, and McGhee, 2001; Wilson et al., 2000), we speculate that some degree of unconscious, automatic gender stereotyping was working behind students' evaluation of the lecturers.

Another possible bias at play is rater-ratee similarity bias. Research has shown that people like and are attracted to those who are similar to them (Newcomb, 1956). Thus, raters favor ratees who are similar by race, gender, personality, or in subtle behaviors (Zalesny and Kirsch, 1989). The two distinct effects, stereotype confirmation bias and rater-ratee similarity bias, could together push male students' ratings of male lecturers up in all measures. In the case of female students, their egalitarian overall evaluation (which agrees with teaching evaluation studies such as Kaschuk (1978) or leadership studies such as Jackson, Engstrom, and Emmers-Sommer (2007)) may reflect a tension between these two biases, which would compete with each other. Further, it has been seen in many studies that that men are more likely than women to be less disposed toward gender egalitarianism (e.g., Glick and Fiske, 1996). In the words of Koch, D'Mello, and Sackett (2015), men's "stronger desire to maintain a segregated occupational system" (p. 130) provides yet another dimension of explanation. Further studies would need to be done to test the validity of these theoretical frameworks for our results.

---

[9] An interesting test of this framework, not part of our study design, would be to have subjects view multiple lecturers, male and female, and then employ a subjective or objective rating scale.

4.4 Recommendations

Administrators who supervise women in STEM fields must learn how to mitigate both implicit and explicit bias, and be aware of the real threat of gender bias evaluations by students. Even if gender bias is only present via an interaction with student gender, since male students are the norm in disciplines like physics, this effect cannot be ignored. Only very recently (Williams and Ceci, 2015) has a psychological study using gender-disguised resumes found that women are significantly favored over men in hypothetical searches for a tenure-track assistant professor in four STEM fields: biology, economics, engineering and psychology. While "efforts to combat formerly widespread sexism in hiring appear to have succeeded" (Williams and Ceci, 2015, p. 5364) in these fields, it is important to note that decades of work and scientific study have contributed to this evidence of social change. Further, because the continued success of a person hired as a new assistant professor is determined by evaluations in the decades to come, this encouraging result does not detract from the need to understand the role of gender in evaluations, particularly by students, of a college professor.

Work by Zastavker et al. (2011) presents the power of gender schemas in engineering education and recommends steps to combat implicit bias, recognize invisible privilege, and avoid negative micro-messaging. Books such as "Faculty Diversity: Problems and Solutions" by Moody (2004) and "Why So Few?" by Hill, Corbett, and St. Rose (2010) are two of the many valuable resources available for faculty and administrators as they seek to overcome the problem of underrepresentation and undervaluation of women in STEM fields. Resources like these are crucial, since the recognition of bias is not enough to mitigate it (Duguid and Thomas-Hunt, 2014; Isaac, Lee, and Carnes, 2009). One must also deliver a counter-stereotyping message, allowing a majority of people to ignore their stereotypical preconceptions. Successful interventions to "break the habit" of race-based prejudice (Devine et al., 2012) and gender-bias (Carnes et al., 2015) have improved academic department climates by creating awareness of prejudice, raising concern about adverse effects of bias, then training participants in multiple bias-reducing strategies.

4.5 Conclusion

In summary, a multiplicity of factors have contributed to the lower fraction of female faculty in STEM fields (Ceci and Williams, 2011; Institute of Medicine et al., 2007; Goulden, Mason, and Frasch, 2011). Our findings suggest that gender bias in evaluations by male students may be one of these factors. Because micro-inequities can add up, resulting in macro-inequities like unbalanced allocation of resources or tenure denials (Rowe, 1990; Valian, 1998), gender bias in student evaluations in the physical sciences has the potential to continue to limit the fraction of female faculty. This would result in negative implications for female STEM students (Carrell, 2009; Stout et al., 2011; Whitten, Foster, and Duncombe, 2003) and STEM professionals alike (Blaha, 2011; Williams, 2015). Moreover, as Whitten (2012) notes, changing the race and gender of the physics community has great ramifications, in terms of "the questions we ask, and the use to which our science is put" (p. 128).

One of the most famous examples of gender-biased evaluation is the Wenneras and Wold (1997) study of the Swedish Medical Research Council (MFR) in which a woman applicant had to be dramatically more productive than a man to earn a postdoctoral fellowship. Less famous is the MRF's response. By 2004, policy changes based on "knowledge on how prejudice influence peer decisions" had eliminated the gender gap in postdoctoral awards (Sandstrom and Hallsten, 2008). Long ago, Tolstoy said that "Everyone thinks of changing the world, but no one thinks of changing himself." But more and more, institutions and individuals are heeding the call to change by "examining our own working practices and attitudes" (Keller, 2011, p. 19).

ACKNOWLEDGMENTS: We are grateful to the reviewers of this manuscript for extremely helpful comments and suggestions. We wish to acknowledge the guidance and diligence of this Journal's Editor during the review process. AG is grateful for support from the Provost's office of Swarthmore College, and to the Mellon Foundation for a New Directions Seed Grant and Sabbatical Award.

**Appendix A:** Professor evaluation questionnaire and item statistics
As described in Methods section above, all questions were evaluated on scale of 1-5. For calculation of overall evaluation score, questions that are starred (*) had their numerical scores reversed, so that a higher number contributed to a more positive overall evaluation.

| | Item | Item Mean | Standard Deviation |
|---|---|---|---|
| 2 | I feel that this professor teaches in a way that really helps students learn | 3.76 | 1.03 |
| 3* | The lecture was difficult to follow. | 3.73 | 1.15 |
| 4 | This professor has a solid grasp of the material being taught. | 3.89 | 1.09 |
| 5* | I was bored by the lecture. | 3.50 | 1.13 |
| 6 | I enjoyed the lecture. | 3.50 | 0.95 |
| 7* | This professor is not a very knowledgeable physicist. | 3.63 | 1.13 |
| 8 | Were I taking a course from this professor, they would be approachable for questions outside of class. | 4.31 | 1.01 |
| 9* | This professor goes too slowly during a class. | 3.23 | 1.21 |
| 10 | The lecture was well-organized. | 3.92 | 1.06 |
| 12* | This professor interacts poorly with the students in class. | 3.98 | 1.04 |
| 13 | This professor is good at handling experimental equipment. | 4.07 | 0.87 |
| 14* | I didn't like this professor. | 3.58 | 1.16 |
| 15 | I think that, overall, the quality of the physics lecture was … | 3.11 | 1.08 |
| 16 | In their role as a physics professor, I'd rate the person I saw as … | 3.23 | 1.09 |
| 17 | If this professor were a candidate for a job teaching at (school name here), I would … | 3.30 | 1.13 |

Figure 1: Interaction between Professor Sex and Student Sex on the Mean Overall Evaluation

Error bars indicate standard error.

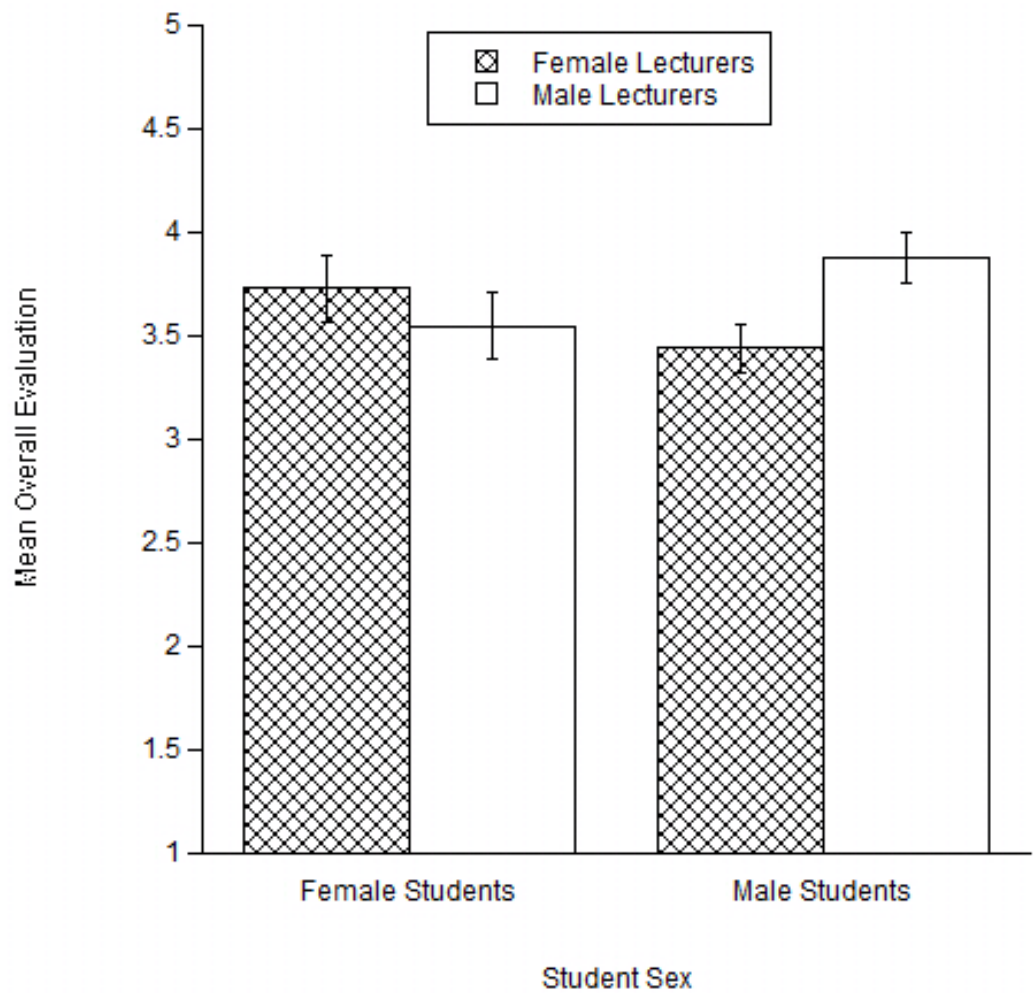

Figure 2: Scissors diagram illustrating direction of interaction between student and lecturer sex, as well as main effect of Institution on the Mean Overall Evaluation. Red circles represent female lecturer condition, blue squares represent male lecturer condition. Institution A is more selective than institution B.

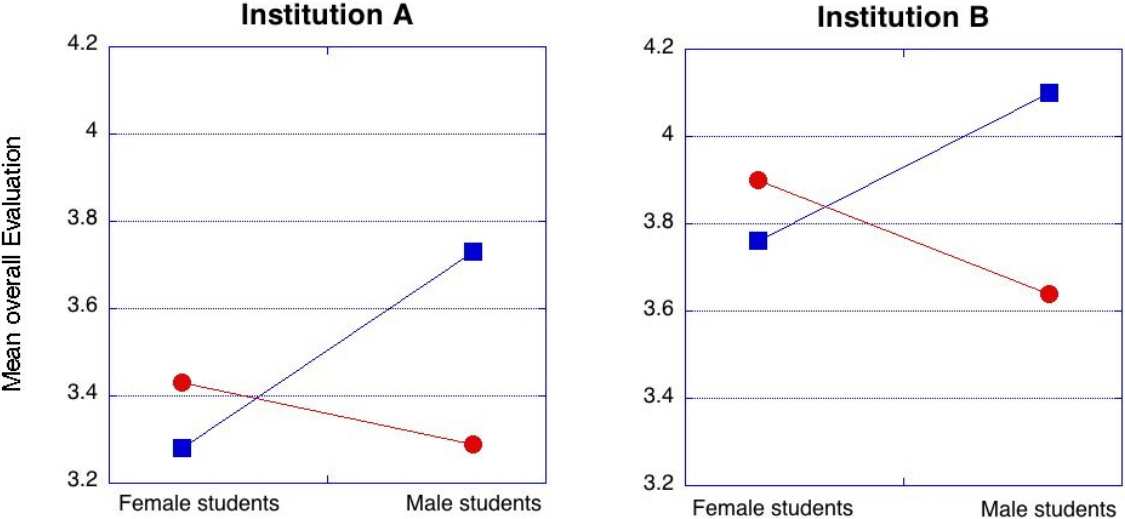

Figure 3: Main Effect of Student Sex on the Mean Evaluation of Scientific Thinking and Hands-on Skills.  Error bars indicate standard error.

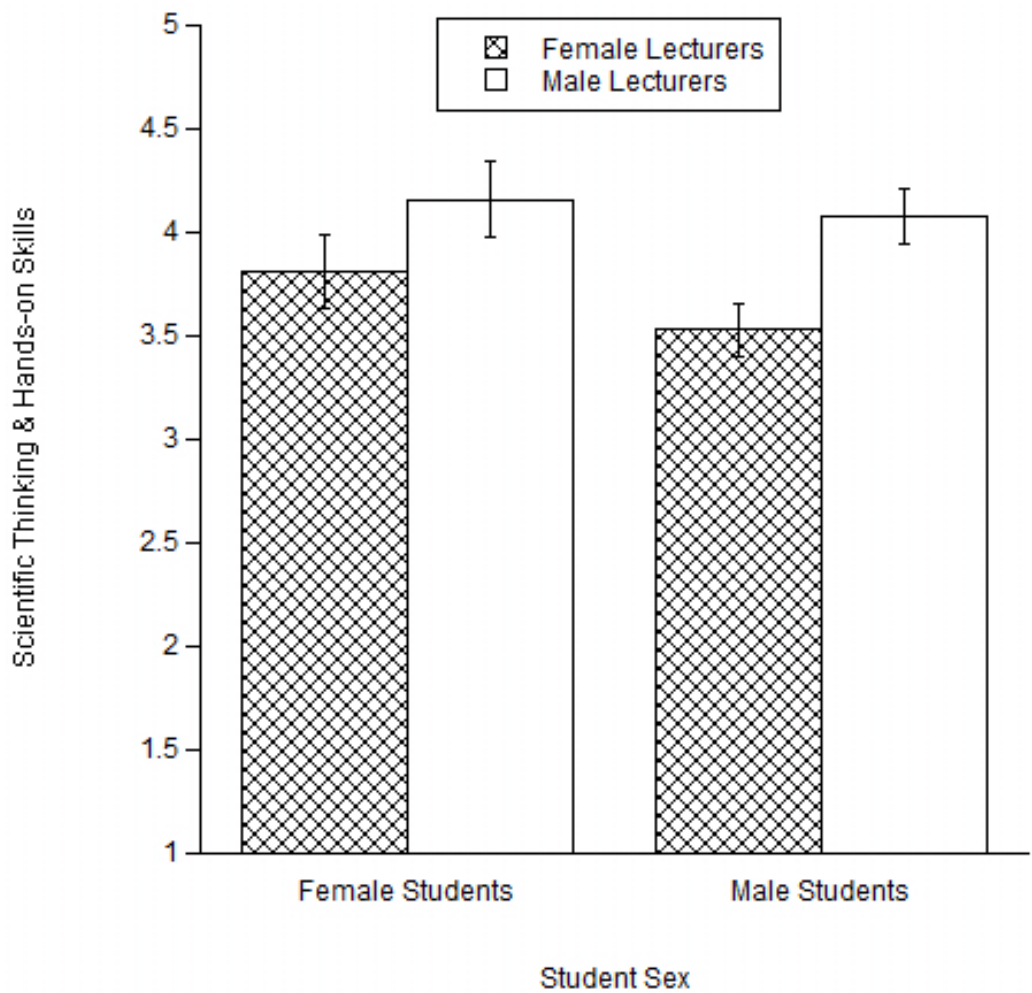

Figure 4: Interaction between Lecturer Sex and Student Sex on the Mean Evaluation of Interpersonal and Communicative Skills.  Error bars indicate standard error.

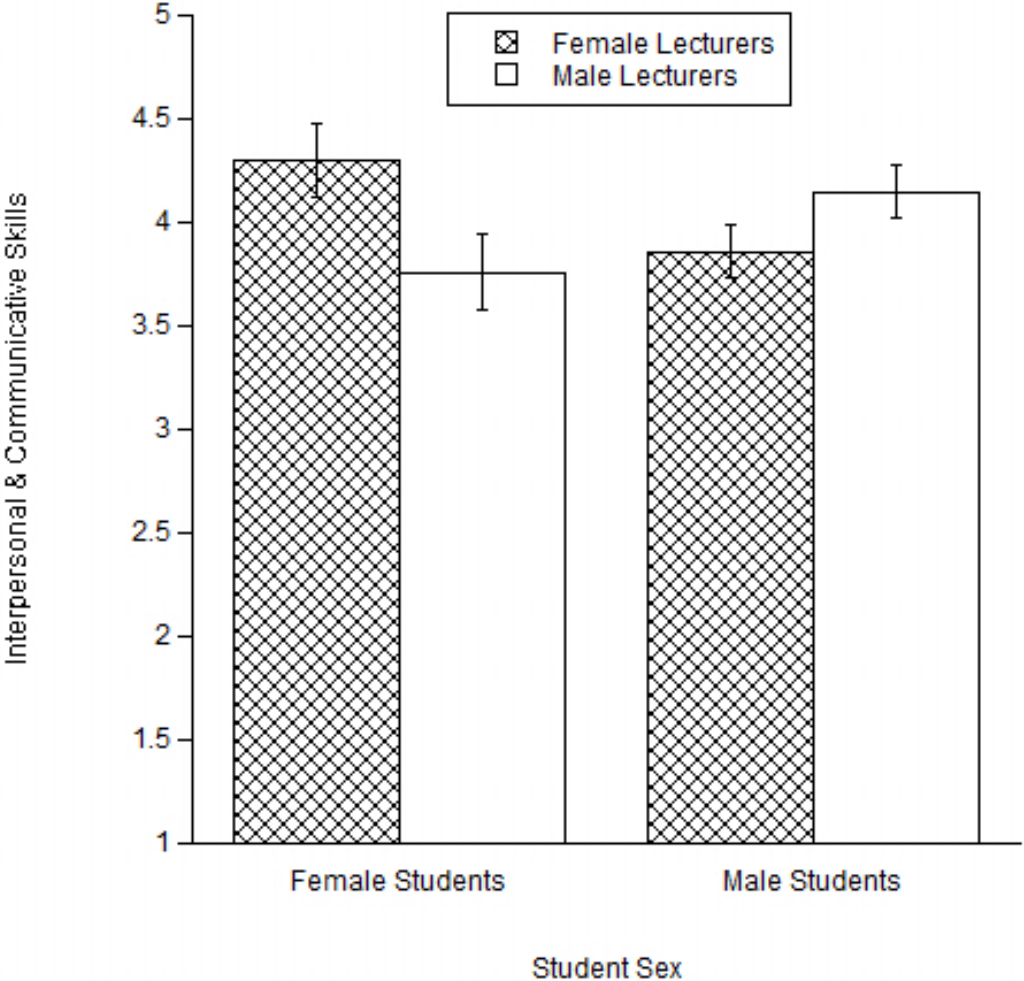

Table 1:   Overall Evaluation

| Female students' rating | | | Male students' rating | | | Overall ANOVA results: |
|---|---|---|---|---|---|---|
| Female lecturers $n = 21$ (Mean & (SD)) | Male lecturers $n = 21$ (Mean & (SD)) | Simple main effect test for the mean differences | Female lecturers $n = 40$ (Mean & (SD)) | Male lecturers $n = 39$ (Mean & (SD)) | Simple main effect test for the mean differences | 1. Main effect of lecturer sex<br>2. Main effect of student sex<br>3. Interaction between lecturer sex & student sex |
| 3.73 (0.70) | 3.55 (0.75) | $F(1, 117) = 0.65$, $p = .423$ | 3.44 (0.71) | 3.88 (0.75) | $F(1, 117) = 7.07$, $p = .008$ | 1. $F(1, 117) = 0.84$, $p = .360$<br>2. $F(1, 117) = 0.01$, $p = .921$<br>3. $F(1, 117) = 4.91$, $p = .029$ |

Note: Some of the questionnaire items were reverse-scored so that higher scores mean more positive evaluation. Reliability of this composite indicated by Cronbach's alpha was .92.

Table 2:   Scientific knowledge and hands-on skills

| Female students' rating | | | Male students' rating | | | Overall ANOVA results: |
|---|---|---|---|---|---|---|
| Female lecturers $n = 21$ (Mean & (SD)) | Male lecturers $n = 21$ (Mean & (SD)) | Simple main effect test for the mean differences | Female lecturers $n = 40$ (Mean & (SD)) | Male lecturers $n = 39$ (Mean & (SD)) | Simple main effect test for the mean differences | 1. Main effect of lecturer sex<br>2. Main effect of student sex<br>3. Interaction between lecturer sex & student sex |
| 3.81 (0.86) | 4.16 (0.80) | $F(1, 117) = 1.90$, $p = .167$ | 3.53 (0.80) | 4.08 (0.83) | $F(1, 117) = 8.91$, $p = .003$ | 1. $F(1, 117) = 8.24$, $p = .005$<br>2. $F(1, 117) = 1.36$, $p = .246$<br>3. $F(1, 117) = 0.42$, $p = .520$ |

This composite includes items: this professor has a solid grasp of the material; this professor is not a very knowledgeable physicist (reversed); and this professor is good at handling equipment. Reliability of this composite indicated by Cronbach's alpha was .76.

Table 3: Interpersonal and communicative skills

| Female students' rating | | | Male students' rating | | | Overall ANOVA results: |
|---|---|---|---|---|---|---|
| Female lecturers $n = 21$ (Mean & (SD)) | Male lecturers $n = 21$ (Mean & (SD)) | Simple main effect test for the mean differences | Female lecturers $n = 40$ (Mean & (SD)) | Male lecturers $n = 39$ (Mean & (SD)) | Simple main effect test for the mean differences | 1. Main effect of lecturer sex<br>2. Main effect of student sex<br>3. Interaction between lecturer sex & student sex |
| 4.30 (0.67) | 3.76 (0.94) | $F(1, 117) = 4.74$, $p = .029$ | 3.86 (0.72) | 4.15 (0.87) | $F(1, 117) = 2.67$, $p = .100$ | 1. $F(1, 117) = 0.63$, $p = .428$<br>2. $F(1, 117) = 0.03$, $p = .867$<br>3. $F(1, 117) = 7.41$, $p = .007$ |

This composite includes items: this professor teaches in a way that helps students learn; this professor would be approachable outside of class; and this professor interacts poorly with the students in class (reversed). Reliability of this composite indicated by Cronbach's alpha was .71.

Table 4: Lecture quality

| Female students' rating | | | Male students' rating | | | Overall ANOVA results: |
|---|---|---|---|---|---|---|
| Female lecturers $n = 21$ (Mean & (SD)) | Male lecturers $n = 21$ (Mean & (SD)) | Simple main effect test for the mean differences | Female lecturers $n = 40$ (Mean & (SD)) | Male lecturers $n = 39$ (Mean & (SD)) | Simple main effect test for the mean differences | 1. Main effect of lecturer sex<br>2. Main effect of student sex<br>3. Interaction between lecturer sex & student sex |
| 3.54 (0.67) | 3.32 (0.83) | $F(1, 117) = 0.89$, $p = .349$ | 3.38 (0.77) | 3.71 (0.76) | $F(1, 117) = 3.53$, $p = .059$ | 1. $F(1, 117) = 0.12$, $p = .733$<br>2. $F(1, 117) = 0.63$, $p = .428$<br>3. $F(1, 117) = 3.50$, $p = .064$ |

This composite includes items: the lecture was difficult to follow (reversed); I was bored by the lecture (reversed); I enjoyed the lecture; this professor goes too slowly (reversed); the lecture was well organized; and overall the quality of lecture was excellent. Reliability of this composite indicated by Cronbach's alpha was .79.